\documentclass[preprintnumbers,amsmath,amssymb]{revtex4}

\usepackage{graphicx}
\usepackage{dcolumn}
\usepackage{bm}
\newcommand{\en}{\end{equation}}
\newcommand{\bea}{\begin{eqnarray}}
\newcommand{\ena}{\end{eqnarray}}
\begin{document}
\title{On a class of scaling FRW cosmological models}
\author{Mauricio Cataldo}
\email{mcataldo@ubiobio.cl} \affiliation{ Departamento de
F\'{\i}sica, Universidad del B\'{\i}o-B\'{\i}o, Avenida Collao 1202,
Casilla 5-C, Concepci\'on, Chile. \\}
\author{Fabiola Arevalo}
\email{farevalo@udec.cl} \affiliation{Departamento de F\'{\i}sica, Universidad de Concepci\'{o}n,\\
Casilla 160-C, Concepci\'{o}n, Chile. \\}
\author{Paul Minning}
\email{pminning@udec.cl} \affiliation{Departamento de F\'{\i}sica, Universidad de Concepci\'{o}n,\\
Casilla 160-C, Concepci\'{o}n, Chile. \\}
\date{\today}
\begin{abstract}
We study Friedmann-Robertson-Walker cosmological models with matter
content composed of two perfect fluids $\rho_1$ and $\rho_2$, with
barotropic pressure densities $p_1/ \rho_1=\omega_1=const$ and $p_2/
\rho_2=\omega_2=const$, where one of the energy densities is given
by $\rho_1=C_1 a^\alpha + C_2 a^\beta$, with $C_1$, $C_2$, $\alpha$
and $\beta$ taking constant values. We solve the field equations by
using the conservation equation without breaking it into two
interacting parts with the help of a coupling interacting term $Q$.
Nevertheless, with the found solution may be associated an
interacting term $Q$, and then a number of cosmological interacting
models studied in the literature correspond to particular cases of
our cosmological model. Specifically those models having constant
coupling parameters $\tilde{\alpha}$, $\tilde{\beta}$ and
interacting terms given by $Q=\tilde{\alpha} H \rho_{_{DM}}$,
$Q=\tilde{\alpha} H \rho_{_{DE}}$, $Q=\tilde{\alpha} H
(\rho_{_{DM}}+ \rho_{_{DE}})$ and $Q=\tilde{\alpha} H
\rho_{_{DM}}+\tilde{\beta} H \rho_{_{DE}}$, where $\rho_{_{DM}}$ and
$\rho_{_{DE}}$ are the energy densities of dark matter and dark
energy respectively. The studied set of solutions contains a class
of cosmological models presenting a scaling behavior at early and at
late times. On the other hand the two-fluid cosmological models
considered in this paper also permit a three fluid interpretation
which is also discussed. In this reinterpretation, for flat
Friedmann-Robertson-Walker cosmologies, the requirement of
positivity of energy densities of the dark matter and dark energy
components allows the state parameter of dark energy to be in the
range $-1.37 \lesssim \omega_{_{DE}}<-1/3$.


\end{abstract}
\maketitle
\section{Introduction}
Recent observational data indicate that our Universe is undergoing
accelerated expansion. The standard modern cosmological models
consider the total energy density of the Universe to be dominated
today by the densities of two components: dark matter (which has an
attractive gravitational effect like usual matter), and dark energy
(a kind of vacuum energy with a negative pressure), which drives the
accelerated expansion~\cite{Lima}. Even more, observational data
seem to indicate that the Universe today may be dominated by an
exotic kind of dark energy which has a very strong negative
pressure, denominated phantom fluid since it violates all energy
conditions~\cite{Caldwell}. The real nature of the dark sector
remains unknown.

Most considered dark energy models present an accelerated expansion
due to the presence of a quintessence (described by a canonical
scalar field), or a phantom field (described by a scalar field with
a negative kinetic term), among others. It appears in this kind of
models that the energy density of dark energy is almost equivalent
to that of the matter in the current or in recent times although
they scale independently during all the cosmic evolution. This
situation is referred to as the ``coincidence problem". To solve the
coincidence problem many attempts have been done. Of special
interest are interacting cosmological models since they may
alleviate or even solve the coincidence problem~\cite{Zimdahl}. We
shall refer to this type of cosmological models below in this
section.

In the framework of General Relativity, for modeling the present
state of the Universe, usually the cosmological models consider two
cosmic fluids as sources for the Einstein field equations: one fluid
for the dark matter sector, where is included the standard visible
matter, and another cosmic fluid for the dark energy sector. Usually
the dark matter is described as a pressureless ideal fluid, while
the dark energy may be described by a cosmological constant, or
perfect fluid or scalar fields~\cite{Sahni}.

The consideration of two fluids in Einstein field equations leads to
\begin{equation}\label{EEQ}
R_{\mu \nu}-\frac{1}{2} R g_{\mu \nu}=\kappa \left(T^1_{\mu
\nu}+T^2_{\mu \nu} \right),
\end{equation}
where $\kappa=8 \pi G$, and $T^1_{\mu \nu}$ and $T^2_{\mu \nu}$ are
the energy--momentum tensors of the two fluids. In a homogeneous and
isotropic FRW universe
\begin{eqnarray}\label{friedmann metric}
ds^2=dt^2-a(t)^2\left (\frac{dr^2}{1-kr^2}-r^2 (d\theta^2+sin^2
\theta d \varphi^2) \right ),
\end{eqnarray}
filled with two fluids $\rho_{_1 }$  and $\rho_{_{2}}$, the
Friedmann equation is given  by
\begin{eqnarray}\label{friedmann equation}
3 H^2= \kappa \left(\rho_{_1 }+\rho_{_{2}}\right)-\frac{k}{a^2}.
\end{eqnarray}
From Eq.~(\ref{EEQ}) it follows that both fluids together form a
system that has a conserved four--momentum, implying that the two
components $\rho_{_1 }$ and $\rho_{_{2}}$ satisfy the following
conservation equation:
\begin{eqnarray}\label{ConEq}
\dot{\rho_{_1 }}+\dot{\rho}_{_{2}}+3 H \left(\rho_{_1 }+\rho_{_{2}
}+p_{_{1}}+p_{_{2}}\right)=0.
\end{eqnarray}
It is clear that the single two-fluid conservation
equation~(\ref{ConEq}) implies that the sum of the two fluids is
conserved. In order to find solutions one may separate this single
two-fluid conservation equation into two conservation equations for
a system of two single fluids:
\begin{eqnarray}\label{ConEq0}
\dot{\rho_{_1 }}+3 H \left(\rho_{_1 }+p_{_{1}}\right)=0,
\nonumber \\
\dot{\rho}_{_{2}}+3 H \left(\rho_{_{2} }+p_{_{2}}\right)=0.
\end{eqnarray}
The physical interpretation is that each of  two fluid components is
conserved, evolving separately according to standard conservation
laws. However, this splitting into two conservation equations is a
condition imposed on the conservation equation~(\ref{ConEq}) and
thus leads to a particular set of solutions to Einstein field
equations. If we apply this picture to the dark sector we conclude
that dark matter and dark energy are conserved separately, and if
the pressures are given by
\begin{eqnarray}\label{5}
p_{_1}(t)=\omega_{_1} \rho{_{_1}}(t), 
p_{_2}(t)=\omega_{_2} \rho{_{_2}}(t),
\end{eqnarray}
where $\omega_{_1}$ and $\omega_{_2}$ are constant parameters, then
the solutions of Eqs.~(\ref{ConEq0}) are given by
\begin{eqnarray}\label{NIF}
\rho_1=\rho_{10} a^{-3(1+\omega_1)}, 
\rho_2=\rho_{20} a^{-3(1+\omega_2)},
\end{eqnarray}
implying that energy density of dark matter has the form
$\rho_{_{DM}}=\rho_{_{DM0}} a^{-3}$.

However, since the real natures of dark matter and dark energy
remain unknown, one can consider more general scenarios where dark
matter and dark energy should not conserve separately and are
coupled to each other. One coupling mechanism can be formally
introduced into the Friedmann equations by defining an interacting
term $Q(t)$ in the following form:
\begin{eqnarray}\label{Q}
\dot{\rho_{_1 }}+3 H \left(\rho_{_1 }+p_{_{1}}\right)=Q(t),
\nonumber \\
\dot{\rho}_{_{2}}+3 H \left(\rho_{_{2} }+p_{_{2}}\right)=-Q(t).
\end{eqnarray}
Note that the case $Q>0$ is interpreted as a transfer of energy from
fluid $\rho{_{_2}}$ to fluid $\rho{_{_1}}$. Then, for the case
$Q<0$, we should have an energy transfer from fluid $\rho{_{_1}}$ to
fluid $\rho{_{_2}}$. In order to find solutions the form of this
phenomenological interacting term $Q$ must be postulated. Usually
one of the following types is considered: $Q=\tilde{\alpha} H
\rho_{_{DM}}$, $Q=\tilde{\alpha} H \rho_{_{DE}}$, $Q=\tilde{\alpha}
H \rho_{_{DM}}+\tilde{\beta} H \rho_{_{DE}}$, $Q=\tilde{\alpha} H
(\rho_{_{DM}}+ \rho_{_{DE}})$, where $\rho_{_{DM}}$ and
$\rho_{_{DE}}$ are the energy densities of dark matter and dark
energy respectively.

As we stated above, the introduction of this type of coupling
between components of the dark sector offers to the standard
cosmology a potential solution to the cosmic coincidence problem by
requiring the ratio of matter and dark energy densities to be stable
against perturbations at late times~\cite{Pavon}. On the other hand,
it also must be noticed that this type of coupling is motivated by
considerations of high energy particle physics~\cite{Farrar}.

It is clear that the assumption of a gravitational coupling between
these two fluids in the form of Eqs.~(\ref{Q}) is too close to the
physical character of the two-fluid conservation equation since the
total energy is conserved always satisfying the conservation
equation~(\ref{ConEq}). Nevertheless, one can look for a general
solution to Einstein equations by using the general properties
encoded into the two-fluid conservation equation by choosing another
more natural way to solve the Friedmann equations~(\ref{friedmann
equation}) and~(\ref{ConEq}), namely by postulating the form of one
of the energy densities based on physical considerations and then
find the whole solution by means of the conservation
equation~(\ref{ConEq}). In this paper this method will be applied.
We shall consider Friedmann-Robertson-Walker (FRW) cosmologies
filled with two ideal barotropic fluids with constant state
parameters, where one of the energy densities is given by a sum of
two powers of the scale factor. For the purposes of this paper, we
shall define a scaling cosmology as one in which the energy
densities $\rho_1$ and $\rho_2$ scale exactly as powers of the scale
factor in the form $\rho_1 \thicksim a^n$ and $\rho_2 \thicksim a^n$
with $n$ constant, leading to the ratio of energy densities of the
form $r=\rho_1/\rho_2= const$. On the other hand, we shall use the
term quintessence not only for a matter content in the form of a
scalar field as is generally used, but also for any kind of
substance having the equation of state $-1<\omega<-1/3$.

The outline of the present paper is as follows: In Section II  we
obtain the general solution for the chosen form of one of energy
densities. In Section III we study some specific two-fluid
interacting cosmologies, considered before in the literature, that
are included in the obtained cosmological models as particular
cases. In Section IV some features of the studied cosmological
models are discussed and a three-fluid interpretation is given.
Finally, Section V presents some concluding remarks.

\section{Field equations for two fluid components}
As we stated in the previous section we shall postulate the form of
one of the energy densities to build a viable cosmological model
which is able to lead to an accelerated expansion and solves or
alleviates the coincidence problem. In order to do this the model
must successfully reproduce some important cosmological scenarios
considered before in standard cosmology. For example, FRW
cosmologies filled with a single fluid $\rho$ with equation of state
$p=\omega \rho$ have an energy density of the form $\rho= \rho_0
a^{-3(\omega+1)}$, and those with two non-interacting fluids, as we
have seen above, have the energy densities given by~(\ref{NIF}).
Another type of FRW cosmologies, which provides us with a way to
approach the coincidence problem, has energy densities which scale
exactly as powers of the scale factor, so they are considered to be
given by $\rho_1=\rho_{10} a^\alpha$ and $\rho_2=\rho_{20} a^\beta$
in order to have $\rho_1/\rho_2 \sim a^{const}$. Just these types of
cosmological scenarios we want to generalize.

The starting point is the assumption that one of energy densities,
say $\rho_1$, depends on the scale factor as
\begin{eqnarray}\label{Rho1}
\rho_1(a)=C_1 a^\alpha+C_2 a^\beta,
\end{eqnarray}
where $C_1$, $C_2$, $\alpha$ and $\beta$ are constants. Then
substituting the expression~(\ref{Rho1}) into the conservation
equation~(\ref{ConEq}) we obtain for the other energy density
\begin{widetext}
\begin{eqnarray}\label{Rho2}
\rho_{_2}(a)=C a^{-3(1+\omega_2)}-\frac{\alpha
+3(1+\omega_1)}{\alpha+3(1+\omega_2)}\, C_1 a^\alpha -
\frac{\beta +3(1+\omega_1)}{\beta+3(1+\omega_2)}\, C_2 a^\beta,
\hspace{2.5cm}
\end{eqnarray}
\end{widetext}
where $C$ is a constant of integration. It is clear that
expressions~(\ref{Rho1}) and~(\ref{Rho2}) may lead to an
asymptotically scaling behavior of energy densities at early and
late times, depending on which term in $\rho_{_1}(a)$ and
$\rho_{_2}(a)$ is dominating at each epoch.

Normalizing expressions~(\ref{Rho1}) and~(\ref{Rho2}) to the present
values $\rho_{10}$ and $\rho_{20}$ (the present value of the scale
factor is normalized as $a=1$) we can express the constants $C_1$
and $C_2$ through the model parameters as

\begin{widetext}
\begin{eqnarray}\label{C1Norm}
C_1=-\frac{\alpha+3(1+\omega_2)}{3(\omega_1-\omega_2)(\beta-\alpha)}
\,
[(\beta+3(1+\omega_1))\rho_{10}+(\beta+3(1+\omega_2))(\rho_{20}-C)],
\end{eqnarray}
\begin{eqnarray}\label{C2Norm}
C_2=\frac{\beta+3(1+\omega_2)}{3(\omega_1-\omega_2)(\beta-\alpha)}
\,
[(\alpha+3(1+\omega_1))\rho_{10}+(\alpha+3(1+\omega_2))(\rho_{20}-C)],
\end{eqnarray}
\end{widetext}
respectively.

Now, by taking into account Eqs.~(\ref{Rho1}) and~(\ref{Rho2}), we
can find the scale factor by solving the Friedmann
equation~(\ref{friedmann equation}) which takes the form
\begin{eqnarray}\label{friedmann equation T}
3 H^2+\frac{k}{a^2}= \kappa \rho_{total},
\end{eqnarray}
where
\begin{eqnarray*}
\rho_{total}=\rho_1+\rho_2=C a^{-3(1+\omega_2)}+ 
\frac{3(\omega_2-\omega_1)}{\alpha+3(1+\omega_2)}\, C_1 a^\alpha +
\frac{3(\omega_2-\omega_1)}{\beta+3(1+\omega_2)}\, C_2 a^\beta.
\end{eqnarray*}
Note that for $C=\rho_{20}$ the energy density~(\ref{Rho2}) takes
the following form:
\begin{eqnarray}
\rho_{_2}(a)=\rho_{20} a^{-3(1+\omega_2)}+ 
\frac{(\alpha +3(1+\omega_1))(\beta
+3(1+\omega_1))\rho_{10}}{3(\omega_1-\omega_2)(\beta-\alpha)}\,
(a^\alpha -a^{\beta}).
\end{eqnarray}
\section{Interacting FRW cosmologies}
As we stated above, in recent years the interacting interpretation
of the conservation equation~(\ref{ConEq}) has received increasing
attention, which is mainly preferred by cosmologists because it
helps to study the coincidence problem by generating two-fluid
cosmological solutions. Clearly, for the solution found in the
previous section, we can introduce into consideration an interacting
term by computing $Q(t)$ with the help of Eq.~(\ref{Q}). Thus, by
substituting Eq.~(\ref{Rho1}) into the first expression of
Eq.~(\ref{Q}) (or Eq.~(\ref{Rho2}) into the second expression of
Eq.~(\ref{Q})), we obtain that the interacting term associated with
our solution~(\ref{Rho1}) and~(\ref{Rho2}) is given by
\begin{eqnarray}\label{Qsolucion}
Q(t)= (\alpha+3(1+\omega_1)) H C_1 a^\alpha + 
(\beta+3(1+\omega_1))H C_2 a^\beta. \hspace{0.315cm}
\end{eqnarray}
Note that we obtain the case of two non-interacting
fluids~(\ref{NIF}) by taking $C_2=0$ and $\alpha+3(1+\omega_1)=0$,
which implies that the interaction term $Q=0$.

Notice that from the general solution~(\ref{Rho1}) and~(\ref{Rho2})
we have that for $C=0$ and
$\alpha=-3\left(1+\frac{1}{2}(\omega_1+\omega_2) \right)$ the energy
densities take the form
\begin{eqnarray*}
\rho_1=\rho_2=C_0 a^{-3-\frac{3}{2} (\omega_1+\omega_2)},
\end{eqnarray*}
where the interaction term is given by
\begin{eqnarray*}
Q=C_0 (\omega_1-\omega_2)a^{-4-\frac{3}{2}(\omega_1+\omega_2)}.
\end{eqnarray*}
Clearly, if $\omega_1=\omega_2$ we obtain the standard case of a FRW
cosmology filled with a single fluid. The interaction in this case
exists only due to the presence of different pressures $p_1=\omega_1
\rho$ and $p_2=\omega_2 \rho$.

Let us now impose on the general expressions for energy
densities~(\ref{Rho1}), (\ref{Rho2}) and the interacting
term~(\ref{Qsolucion}) some types of specific interactions
considered in the literature. In general the considered interacting
terms are functions of the energy densities multiplied by a function
with units of inverse of time (generally $H(t)=\dot{a}/a)$. As we
quoted in the introduction some often considered types are:
$Q=\tilde{\alpha} H \rho_{_{DM}}$, $Q=\tilde{\alpha} H
\rho_{_{DE}}$, $Q=\tilde{\alpha} H (\rho_{_{DM}}+ \rho_{_{DE}})$,
$Q=\tilde{\alpha} H \rho_{_{DM}}+\tilde{\beta} H \rho_{_{DE}}$,
where $\rho_{_{DM}}$ and $\rho_{_{DE}}$ are the energy density of
dark matter and dark energy respectively. We will therefore
consider, in the rest of this section, these types of interactions
in order to link them associated with our solution interacting
term~(\ref{Qsolucion}).

\subsection{Imposing an interacting term proportional to the Hubble parameter and to one of the energy densities}
Let us first consider that the interacting term $Q(t)$ is given by
\begin{eqnarray}\label{QH1}
Q(t)= \tilde{\alpha} H \rho_1,
\end{eqnarray}
where $\tilde{\alpha}$ is a dimensionless constant parameter. Thus,
imposing on the obtained cosmological solution the
condition~(\ref{QH1}), we obtain
\begin{eqnarray}\label{QQ1}
Q(t)= \tilde{\alpha} H \rho_1=\tilde{\alpha} H \left(C_1
a^\alpha+C_2 a^\beta  \right).
\end{eqnarray}
Comparing Eqs.~(\ref{Qsolucion}) and~(\ref{QQ1}) we obtain the
following constraints:
\begin{eqnarray}\label{CQH1}
\tilde{\alpha}=\alpha+3(1+\omega_{{1}}), 
\tilde{\alpha}=\beta+3(1+\omega_{{1}}),
\end{eqnarray}
implying that
\begin{eqnarray}\label{alphaCQH1}
\alpha=\beta=\tilde{\alpha}-3(1+\omega_{{1}}) .
\end{eqnarray}
This implies that the energy densities now may be written in the
following $5$-parametric form:
\begin{eqnarray}\label{alphaRHO1}
\rho_1(a)=\rho_{10}\, a^{\tilde{\alpha}-3(1+\omega_{{1}})},
\hspace{2.7cm}\\
\rho_2(a)=\rho_{20} a^{-3(1+\omega_2)}+ 
\label{alphaRHO2} \frac{\tilde{\alpha}
\rho_{10}}{\tilde{\alpha}+3(\omega_2-\omega_1)} \,
\left(a^{-3(1+\omega_2)} -a^{\tilde{\alpha}-3(1+\omega_1)}\right).
\end{eqnarray}
Let us now suppose that the interacting term is given by
\begin{eqnarray}\label{QH2}
Q(t)= \tilde{\beta} H \rho_2,
\end{eqnarray}
with $\tilde{\beta}$ a dimensionless constant parameter. Thus, by
taking into account Eqs.~(\ref{Rho2}), (\ref{Qsolucion})
and~(\ref{QH2}), we obtain the following condition:
\begin{eqnarray}
(\alpha+3(1+\omega_1)) C_1 a^\alpha + (\beta+3(1+\omega_1))C_2
a^\beta= 
\tilde{\beta} \left( C a^{-3(1+\omega_2)}-\frac{\alpha
+3(1+\omega_1)}{\alpha+3(1+\omega_2)}\, C_1 a^\alpha \right.-
\left. \frac{\beta +3(1+\omega_1)}{\beta+3(1+\omega_2)}\, C_2
a^\beta \right). 
\end{eqnarray}
It is clear that we must put $C=0$ in order to have a
self-consistent interacting solution and the following constraints
must be imposed:
\begin{eqnarray}\label{CQH2}
\tilde{\beta}=-\alpha-3(1+\omega_{{2}}), 
\tilde{\beta}=-\beta-3(1+\omega_{{2}}),
\end{eqnarray}
implying that
\begin{eqnarray}\label{alphaCQH2}
\alpha=\beta=-\tilde{\beta}-3(1+\omega_{{2}}) .
\end{eqnarray}
Thus from Eqs.~(\ref{Rho1}) and~(\ref{Rho2}) we conclude that these
interacting scenarios may be written as
\begin{eqnarray}\label{Rho22}
\rho_1(a)=\tilde{\rho}_{10} a^{-\tilde{\beta}-3(1+\omega_2)}, \hspace{2.3cm} \\
\rho_2(a)=\frac{3(\omega_1-\omega_2)-\tilde{\beta}}{\tilde{\beta}}
\, \tilde{\rho}_{10} a^{-\tilde{\beta}-3(1+\omega_2)},
\label{Rho22A}
\end{eqnarray}
where $\tilde{\rho}_{10}=C_1+C_2$. It is clear that in this case we
have a scaling behavior of energy densities where
\begin{eqnarray}
r=\frac{\rho_2}{\rho_1}=\frac{3(\omega_1-\omega_2)-\tilde{\beta}}{\tilde{\beta}}.
\end{eqnarray}
This ratio is positive for $0<\tilde{\beta}<3(\omega_1-\omega_2)$
when $\omega_1>\omega_2$, and $3(\omega_1-\omega_2)<\tilde{\beta}<0$
when $\omega_1<\omega_2$. For $\tilde{\beta}=3(\omega_1-\omega_2)$
we obtain the standard FRW solution for a single ideal, and for
$\omega_1=\omega_2$ we have that $\rho_1=-\rho_2$, thus obtaining a
vacuum FRW cosmology.

Nevertheless, notice that the $4$-parametric solution~(\ref{Rho22})
and~(\ref{Rho22A}) is a particular case of Eqs.~(\ref{alphaRHO1})
and~(\ref{alphaRHO2}). Effectively, Eqs.~(\ref{Rho22})
and~(\ref{Rho22A}) are obtained if we put
$\rho_{20}=-\frac{\tilde{\alpha}
\rho_{10}}{\tilde{\alpha}+3(\omega_2-\omega_1)}$ and
$\tilde{\alpha}=-\tilde{\beta}+3(\omega_1-\omega_2)$ into
Eqs.~(\ref{alphaRHO1}) and~(\ref{alphaRHO2}). Therefore, since the
coupling~(\ref{QH2}) is a particular case of Eq.~(\ref{QH1}), the
general solution for a coupling proportional to the Hubble parameter
and to one of the energy densities is given by
Eqs.~(\ref{alphaRHO1}) and~(\ref{alphaRHO2}).

Now we shall consider some particular cases of interacting
cosmological scenarios described by Eqs.~(\ref{QH1}),
(\ref{alphaRHO1}) and~(\ref{alphaRHO2}) in order to confront them
with those discussed in the literature.

\subsubsection{Interacting term proportional to dark matter energy density}
Let us first suppose that $\rho_1$ describes the energy density of
the dark matter component $\rho_{_{DM}}$. This implies that we must
put $\omega_1=0$. In this case the interacting term~(\ref{QH1}) may
be written as $Q(t)=\tilde{\alpha}H\rho_{_{DM}}$ and the energy
densities~(\ref{alphaRHO1}) and~(\ref{alphaRHO2}) as follows:
\begin{eqnarray}\label{alphaRHO1DM}
\rho_{_{DM}}:=\rho_1(a)=\rho_{_{DM0}}\, a^{\tilde{\alpha}-3 }, \hspace{2.7cm}\\
\rho_{_{DE}}:=\rho_2(a)=\rho_{_{DE0}} a^{-3(1+\omega_2)}+
\label{alphaRHO2DM} \frac{\tilde{\alpha}
\rho_{_{DM0}}}{\tilde{\alpha}+3\omega_2} \, \left(a^{-3(1+\omega_2)}
-a^{\tilde{\alpha}-3}\right),
\end{eqnarray}
where $\rho_{_{DM0}}$ and $\rho_{_{DE0}}$ are positive constants.
For $-1\leq \omega_2 < -1/3$ the dark energy $\rho_2$ may be
interpreted as quintessence, and for $\omega_2<-1$ as phantom
matter. Clearly, due to the interaction~(\ref{QH1}),
Eq.~(\ref{alphaRHO1DM}) is a deviation from the standard behavior of
the matter component $\rho_{_{DM}} \sim a^{-3}$, which implies that
matter is conserved separately from $\rho_2$. Some aspects of this
kind of interacting scenarios were considered and discussed by
authors of Ref.~\cite{IntDM}. Let us now enumerate other relevant
features of these cosmologies:

By supposing that the dark matter energy density decreases with the
expansion we see from Eq.~(\ref{alphaRHO1DM}) that we must require
that $\tilde{\alpha}-3<0$. On the other hand, from now on we will be
considering only positive energy densities. Thus, in order for the
energy densities~(\ref{alphaRHO1DM}) and~(\ref{alphaRHO2DM}) to
correspond to positive densities during all evolution, we must
require that $\frac{\tilde{\alpha}}{\tilde{\alpha}+3\omega_2}<0$ and
$\rho_{_{DE0}}+\frac{\tilde{\alpha} \rho_{_{DM0}}}{\tilde{\alpha}+3
\omega_2} >0 $. The condition
$\frac{\tilde{\alpha}}{\tilde{\alpha}+3\omega_2}<0$ implies that
$\tilde{\alpha}>0$ and $\tilde{\alpha}+3 \omega_2<0$ or that
$\tilde{\alpha}<0$ and $\tilde{\alpha}+3 \omega_2>0$. Any other
combination for $\tilde{\alpha}$ and $\omega_2$ will necessarily
imply that the energy density of dark energy becomes negative for
some value of the scale factor $a$ during the cosmic evolution.

Note that if $\tilde{\alpha}<0$ and $\tilde{\alpha}+3 \omega_2>0$ we
have that $\omega_2 >0$ and then in this case the consideration of
dark energy is excluded. There exists the possibility of considering
$\omega_2<-1/3$ for constraints $\tilde{\alpha}>0$ and
$\tilde{\alpha}+3 \omega_2<0$ which imply that $\omega_2<0$. In this
case the coupling parameter $\tilde{\alpha}$ is constrained to be
$0<\tilde{\alpha} <3$, implying that in these scenarios the energy
is transferred from dark energy to dark matter. At early times we
have that the dark energy behaves as $\rho_2 \simeq
-\frac{\tilde{\alpha}\rho_{_{DM0}}}{\tilde{\alpha}+3 \omega_2} \,
a^{-3+\tilde{\alpha}}$ and then $r=\rho_{_{DM}}/\rho_{_{DE}}\simeq
-(\tilde{\alpha}+3\omega_2)/\tilde{\alpha}$, and for late times $r
\rightarrow 0$ implying that the dark energy dominates over dark
matter. For this model one finds that the equilibrium between dark
matter and dark energy, i.e. $r(a_{eq})=1$, corresponds to
\begin{eqnarray*}
a_{eq}= \left( \frac{\tilde{\alpha}}{2 \tilde{\alpha}+3 \omega_2}+
\frac{\tilde{\alpha}+3 \omega_2}{2\tilde{\alpha}+3 \omega_2}
\frac{\rho_{_{DE0}}}{\rho_{_{DM0}}}
\right)^{\frac{1}{\tilde{\alpha}+3 \omega_2}},
\end{eqnarray*}
while the accelerated expansion begins at
\begin{eqnarray*}
a_{ac}= \left(\frac{\tilde{\alpha}(1+3 \omega_2)}{3
\omega_2(\tilde{\alpha}-1)}+ \frac{(1+3 \omega_2)(\tilde{\alpha}+3
\omega_2)}{3 \omega_2(\tilde{\alpha}-1)}
\frac{\rho_{_{DE0}}}{\rho_{_{DM0}}}
\right)^{\frac{1}{\tilde{\alpha}+3 \omega_2}}.
\end{eqnarray*}
Clearly these two expressions do not coincide. It can be shown that
for $\omega_2<-2/3$ we have $a_{ac}<a_{eq}$ implying that we can
have scenarios where the universe already has entered into an
accelerated expansion while the dark matter component is still
dominating. For $-2/3<\omega_2<-1/3$ we have that $a_{ac} > a_{eq}$
and then the acceleration begins when the dark energy component
already dominates in the Universe. It is interesting to remark that
this kind of behavior also is observed in power law cosmologies
studied in Ref.~\cite{Cataldo}

Let us note that, in this case, in order to have a transfer of
energy from the dark matter component to dark energy we need to
require $\tilde{\alpha} <0$, then having a negative dark energy at
early times.

\subsubsection{Interacting term proportional to dark energy density}
It is interesting to note that the considered energy
densities~(\ref{alphaRHO1}) and~(\ref{alphaRHO2}) are not symmetric
for the state parameter interchange $(\omega_1=0, \omega_2=\omega)
\leftrightharpoons (\omega_1=\omega, \omega_2=0)$, where $\omega$ is
a free parameter, as well as for the general solution~(\ref{Rho1})
and~(\ref{Rho2}). So for scenarios containing interacting dark
energy and dark matter with $\omega_2=0$ the energy densities are
given by
\begin{eqnarray}\label{alphaRHO1A}
\rho_{_{DE}}:=\rho_1(a)=\rho_{_{DE0}}\, a^{\tilde{\alpha}-3(1+\omega_{{1}})}, \hspace{2.7cm}\\
\label{alphaRHO2A}
\rho_{_{DM}}:=\rho_2(a)=\rho_{_{DM0}} a^{-3 }+ 
\frac{\tilde{\alpha} \rho_{_{DE0}}}{\tilde{\alpha}-3\omega_1} \,
\left(a^{-3 } -a^{\tilde{\alpha}-3(1+\omega_1)}\right).
\end{eqnarray}
The interpretation of this qualitatively different interacting
cosmology is direct: $\rho_2$ is now the energy density of dark
matter while $\rho_1$ is the energy density of dark energy for
$\omega_1<-1/3$. Thus from Eq.~(\ref{QH1}) we conclude that now the
interacting term is proportional to dark energy density, i.e.
$Q(t)=\tilde{\alpha} H \rho_{_{DE}}$. This solution is a particular
solution of the interacting scenarios discussed in Ref.~\cite{He}
(see Sec. IIA).

Now we shall enumerate some relevant features of these cosmologies.
Note that in order to have a positive energy density for dark matter
during all evolution we must require that
$\tilde{\alpha}/(\tilde{\alpha}-3 \omega_1)<0$ and
$\rho_{_{DM0}}+(\tilde{\alpha}/(\tilde{\alpha}-3 \omega_1))
\rho_{_{DE0}}>~0$. The condition $\tilde{\alpha}/(\tilde{\alpha}-3
\omega_1)<0$ implies that $\tilde{\alpha}>0$ and $\tilde{\alpha}-3
\omega_1<0$ or that $\tilde{\alpha} <0$ and $\tilde{\alpha}-3
\omega_1>0$. On the other hand, by supposing that the energy density
of dark matter~(\ref{alphaRHO2A}) decreases with the scale factor we
must require that $\tilde{\alpha}-3(1+\omega_1)<0$.

For $\tilde{\alpha}>0$ and $\tilde{\alpha}-3 \omega_1<0$ we conclude
that $\omega_1>0$ and then the consideration of dark energy is
excluded in this case. On the other hand, for $\tilde{\alpha}<0$ and
$\tilde{\alpha}-3 \omega_1>~0$ we conclude that $\omega_1<0$,
implying finally that the constraint on the coupling parameter
$\tilde{\alpha}$ must be $3 \omega_1< \tilde{\alpha} \leq
3(1+\omega_1)$. In this case at early times the dark matter
component dominates over dark energy and for late times we have a
scaling behavior for energy densities since
$r=\rho_{_{DM}}/\rho_{_{DE}}\simeq -\tilde{\alpha}/(\tilde{\alpha}-3
\omega_1)$. In such scenarios the transfer of energy goes from the
dark energy to the dark matter component.

In general for these models we have that the equilibrium between
dark matter and dark energy corresponds to
\begin{eqnarray*}
a_{eq}= \left( \frac{\tilde{\alpha}}{2 \tilde{\alpha}-3 \omega_1}+
\frac{3 \omega_1-\tilde{\alpha}}{3 \omega_1-2\tilde{\alpha}}
\frac{\rho_{_{DM0}}}{\rho_{_{DE0}}}
\right)^{\frac{1}{\tilde{\alpha}-3 \omega_1}},
\end{eqnarray*}
while the accelerated expansion begins at
\begin{eqnarray*}
a_{ac}= \left(\frac{\tilde{\alpha}}{3
\omega_1(1+3\omega_1-\tilde{\alpha})}+ \frac{(\tilde{\alpha}-3
\omega_1)}{3 \omega_1(1+3\omega_1-\tilde{\alpha})}
\frac{\rho_{_{DM0}}}{\rho_{_{DE0}}}
\right)^{\frac{1}{\tilde{\alpha}-3 \omega_1}}.
\end{eqnarray*}
As before we can have scenarios where the universe already has
entered into an accelerated expansion while the dark matter
component is still dominating.

Lastly, notice that in this case in order to have a transfer of
energy from dark matter to dark energy we must require a negative
dark energy during all cosmic evolution, or $\tilde{\alpha}
>0$ with a negative dark matter at early or late times.

\subsection{Imposing an interacting term proportional to the Hubble parameter and to a linear combination of the energy densities}
Let us now consider the interacting term given by
\begin{eqnarray}\label{Q12}
Q(t)=\tilde{\alpha} H \rho_1+\tilde{\beta} H \rho_2.
\end{eqnarray}
Thus from Eqs.~(\ref{Rho1}), (\ref{Rho2}),~(\ref{Qsolucion})
and~(\ref{Q12}) we firstly conclude that we must put $C=0$. Thus the
constraints on the model parameters are
\begin{eqnarray}\label{parameters12}
\tilde{\alpha}={\frac {\tilde{\beta}\, \left(
\alpha+3(1+\,\omega_{{1}}) \right) }{\alpha+3(1+
\,\omega_{{2}})}}+\alpha+3(1+\,\omega_{{1}}) , 
\tilde{\alpha}={\frac {\tilde{\beta}\, \left(
\beta+3(1+\,\omega_{{1}}) \right) }{\beta+3(1+\,
\omega_{{2}})}}+\beta+3(1+\,\omega_{{1}}).
\end{eqnarray}
Note that these conditions include the cases~(\ref{QH1})
and~(\ref{QH2}) considered above since for $\tilde{\alpha}=0$ we
obtain the conditions~(\ref{CQH2}), and for $\tilde{\beta}=0$ we
obtain the conditions~(\ref{CQH1}). However note that in the
procedure below we assume that $\tilde{\beta} \neq 0$ in order to
self-consistently solve the now treated problem.

In  the following we shall express the powers of the scale factor
$\alpha$ and $\beta$ as functions of the model parameters
$\omega_{{1}}$, $\omega_{{2}}$, $\tilde{\alpha}$ and
$\tilde{\beta}$. Thus from Eqs.~(\ref{parameters12}) we conclude
that the powers of the scale factors are constrained to be given by
the expressions
\begin{eqnarray}\label{Bar}
\alpha=p + \frac{1}{2}\, \epsilon_{_\alpha} \sqrt{q}, 
\beta=p +\frac{1}{2} \epsilon_{_\beta}  \, \sqrt{q},
\end{eqnarray}
where $\epsilon_{_\alpha}=\pm 1$, $\epsilon_{_\beta}=\pm 1$ and
\begin{eqnarray}
q=\left( {\tilde{\alpha}}-{\tilde{\beta}} \right) ^{2}+9\, \left(
\omega_{{1}}-\omega_{{2}} \right) ^{2}+ 6\left(
{\tilde{\alpha}}+{\tilde{\beta}} \right)  \left(
\omega_{{2}}-\omega_{{1}} \right), 
p=-\frac{1}{2}\,\left(\tilde{\alpha}-\tilde{\beta}\right)-\frac{3}{2}\,\left(\omega_{{1}}+\omega_{{2}}+2
\right). \hspace{2.3cm}
\end{eqnarray}
Thus the interacting cosmological scenarios take the following
forms:
\begin{eqnarray}\label{BS}
\rho_1(a)= C_1 a^{p +\frac{1}{2} \epsilon_{_\alpha} \sqrt{q}}+C_2 a^{p +\frac{1}{2} \epsilon_{_\beta} \sqrt{q}} \hspace{2.05cm}  \\
\label{BSA} \rho_2(a)= -\frac{p + \frac{1}{2} \epsilon_{_\alpha}
\sqrt{q}+3(1+\omega_1)}{p + \frac{1}{2} \epsilon_{_\alpha}
\sqrt{q}+3(1+\omega_2)} \,C_1 a^{p +\frac{1}{2} \epsilon_{_\alpha}
\sqrt{q}}- 
\frac{p + \frac{1}{2} \epsilon_{_\beta}
\sqrt{q}+3(1+\omega_1)}{p + \frac{1}{2} \epsilon_{_\beta}
\sqrt{q}+3(1+\omega_2)} \, C_2  a^{p + \frac{1}{2} \epsilon_{_\beta}
\sqrt{q}}. \hspace{0.34cm}
\end{eqnarray}
Clearly we can have scenarios where $\alpha \neq \beta$ by taking
$\epsilon_{_\alpha} =-\epsilon_{_\beta}$, and scenarios where
$\alpha=\beta$ by taking $\epsilon_{_\alpha} =\epsilon_{_\beta}$ or
by requiring $q=0$. This latter condition will imply that
$\tilde{\alpha}=3(\omega_1-\omega_2)+\tilde{\beta} \pm 2 \sqrt{3 \,
\tilde{\beta} (\omega_1-\omega_2)}$ and then the energy densities
take the following particular form:
\begin{eqnarray}
\rho_1(a)=D a^{-3(1+\omega_2) \pm
\sqrt{3\tilde{\beta}(\omega_1-\omega_2)}}, \hspace{2.15cm}\nonumber \\
\rho_2(a)=\mp \frac{D\left(3(\omega_1-\omega_2) \pm
\sqrt{3\tilde{\beta}(\omega_1-\omega_2)}\right)}{\sqrt{3\tilde{\beta}(\omega_1-\omega_2)}}
\times 
a^{-3(1+\omega_2) \pm
\sqrt{3\tilde{\beta}(\omega_1-\omega_2)}},
\end{eqnarray}
where $D$ is a new constant.

This kind of solutions was considered by authors of
Ref.~\cite{Barrow}. Our solution takes the form of the solution
discussed in Ref.~\cite{Barrow} by means of
$\omega_1=\Gamma_{_B}-1$, $\omega_2=\gamma_{_B}-1$,
$\tilde{\alpha}=-\beta_{_B}$ and $\tilde{\beta}=\alpha_{_B}$, where
all parameters with subscript $B$ denote the parameters used by
Barrow and Clifton in cited references. However, it must be noticed
that in this case we can not have a scaling behavior for the energy
densities $\rho_1$ and $\rho_2$ of the form $\rho_2=\lambda (C_3
a^u+C_4 a^v)=\lambda \rho_1$, with $\lambda$, $C_3$, $C_4$, $u$ and
$v$ constants. This can be seen by direct integration of
Eqs.~(\ref{Q}) with the interacting term~(\ref{Q12}) by imposing the
condition $\rho_2=\lambda \rho_1$ with $\lambda$ constant. In this
case the general self-consistent solution is given by
$\rho_1=\rho_{10}
a^{-3(1+\frac{\omega_1+\lambda\omega_2}{1+\lambda})}$. This scaling
solution also can be directly obtained from the general
solution~(\ref{Rho1}) and~(\ref{Rho2}) by imposing the same
condition $\rho_2=\lambda \rho_1$.

Another scenario considered in interacting cosmologies is that
defined by an interacting term of the form
$Q(t)=\tilde{\alpha}H(\rho_1+\rho_2)$. These cosmologies are a
particular case of Eqs.~(\ref{Q12}) and~(\ref{parameters12}) and may
be obtained from Eqs.~(\ref{Bar})-(\ref{BSA}) by putting
$\tilde{\alpha}=\tilde{\beta}$. In general this type of interacting
terms has been discussed in the framework of scalar
fields~\cite{Chimento}.

Lastly notice that the solution~(\ref{BS}) and~(\ref{BSA}) does not
include as a particular case the whole class of interacting
cosmologies given by~(\ref{QH1}) and described by
Eqs.~(\ref{alphaRHO1}) and~(\ref{alphaRHO2}). This is due to the
fact that for $\tilde{\beta}=0$ we must impose the extra conditions
$\beta+3(1+ \,\omega_{{2}})=1$ and $\alpha+3(1+ \,\omega_{{2}})=1$
as we can see from Eqs.~(\ref{parameters12}), thus obtaining only
particular solutions of Eqs.~(\ref{alphaRHO1})
and~(\ref{alphaRHO2}). For the case $\tilde{\alpha}=0$
Eqs.~(\ref{parameters12}) do not impose any extra condition on the
model parameters. Thus the whole class of interacting
cosmologies~(\ref{Rho22}) and~(\ref{Rho22A}) is included in
Eqs.~(\ref{BS}) and~(\ref{BSA}) as a particular case.

\section{Some features of the general solution}
Now we shall study the cosmological scenarios derived in Sec. II by
the assumptions that one component of the universe obeys an equation
of the form~(\ref{Rho1}) and that pressures obey barotropic
equations of state with constant state parameters. As we have seen
cosmologies described by Eqs.~(\ref{Rho1}) and~(\ref{Rho2}), with
arbitrary $\alpha$ and $\beta$, generalize the previous interacting
cosmological scenarios considered in Sec. III. So clearly the found
analytical cosmological models provide us with extensions of the
analysis of cosmologies containing dark matter and dark energy.

\subsection{Cosmologies with $\omega_1=0$}
Let us first consider generalizations of the model described by
Eqs.~(\ref{alphaRHO1DM}) and~(\ref{alphaRHO2DM}). This implies that
we must put $\omega_1=0$ into Eqs.~(\ref{Rho1}) and~(\ref{Rho2}).
Note that for $C_2=0$, $C_1=\rho_{_{DM0}}$,
$C=\rho_{_{DE0}}+\frac{\tilde{\alpha}\rho_{_{DM0}}}{\tilde{\alpha}+3
\omega_2}$ and $\alpha=\tilde{\alpha}-3$ (or $C_1=0$,
$C_2=\rho_{_{DM0}}$,
$C=\rho_{_{DE0}}+\frac{\tilde{\alpha}\rho_{_{DM0}}}{\tilde{\alpha}+3
\omega_2}$ and $\beta=\tilde{\alpha}-3$) we obtain the same
cosmologies described by Eqs.~(\ref{alphaRHO1DM})
and~(\ref{alphaRHO2DM}). Thus the extension of the analysis is
obtained for $C_1 \neq 0$ and $C_2 \neq 0$ simultaneously. In this
case Eq.~(\ref{Rho1}) describes the behavior of the dark matter and
Eq.~(\ref{Rho2}) describes the behavior of the dark energy. By
taking into account Eqs.~(\ref{C1Norm}) and~(\ref{C2Norm}) (with
$\rho_{10}=\rho_{_{DM0}}$ and $\rho_{20}=\rho_{_{DE0}}$) the dark
energy component now is given by
\begin{widetext}
\begin{eqnarray}\label{Rho2DE}
\rho_{_{DE}}= C a^{-3(1+\omega_2)}+ \hspace{12.215cm} \nonumber \\
 \frac{(\alpha +3)(\beta+3)\rho_{_{DM0}}}{3
\omega_2(\alpha -\beta)}\, \left(a^\alpha-a^{\beta}\right) +
\frac{\rho_{_{DE0}}-C}{3
\omega_2(\alpha-\beta)}\left[(\alpha+3)(\beta+3(1+\omega_2))a^\alpha-(\beta+3)(\alpha+3(1+\omega_2))a^\beta\right].
\end{eqnarray}
\end{widetext}

Clearly, in general, there are values of the model parameters which
lead to negative values of the dark matter and/or the dark energy
component. These energies can be either positive during all cosmic
evolution or negative at the beginning or at the end of the
expansion of the Universe. It must be noticed that negative values
of dark energy could in principle be allowed if, for example, this
type of energy is a manifestation of modified
gravity~\cite{Amendola}.

In the following for the sake of simplicity we shall consider the
case $C=\rho_{_{DE0}}$. Clearly this particular form of the general
solution is still a generalization of Eqs.~(\ref{alphaRHO1DM})
and~(\ref{alphaRHO2DM}), which can be obtained by putting
$\alpha=\tilde{\alpha}-3$ and $\beta=-3(1+\omega_2)$. By requiring
the positivity of the dark matter the constraints $C_1
>0$ and $C_2>0$ must be fulfilled. On the other hand, in order for
dark matter to decrease during all cosmic evolution we must require
that $\alpha<0$ and $\beta<0$ and, since $\omega_2$ is the state
parameter of the dark energy, we shall constrain it to
$\omega_2<-1/3$. Without any loss of generality we can consider
$\alpha>\beta$. Thus we have that $\rho_{_{DM}}
>0$ during all cosmic evolution if i) $-2<\beta<\alpha<0$,
$-\frac{\alpha+3}{3}<\omega_2<-\frac{\beta+3}{3}$ and $-2<\alpha<0$,
$-3<\beta<-2$ (in this case $-\frac{\alpha+3}{3}<\omega_2<-1/3$) or
ii) $-3<\alpha<0$ and $\beta<-3$ for any $\omega_2<-1/3$.

Now, by taking into account Eq.~(\ref{Qsolucion}), we conclude that
for the case i) we have energy transfer from dark energy to dark
matter ($Q>0$), while for the case ii) we have that at early times
the energy is being transferred from dark matter to dark energy
($Q<0$) and at late times we have a transfer of energy from dark
energy to dark matter ($Q>0$).

Lastly notice that, in general, independent of which term among
$a^{\alpha}$, $a^{\beta}$ and $a^{-3(1+\omega_2)}$ is dominating at
early and at late times, we may have a scaling behavior for energy
densities at early and late epochs. However, we are interested in
such a behavior of the considered set of solutions at late times.
Without any lost of generality we can suppose that $\beta< \alpha
<0$. Thus, in this case the dark matter component will behave as
$\rho_{_{DM}} \sim C_1 a^\alpha$. For the dark energy component we
can have $\rho_{_{DE}} \sim C a^{-3(1+\omega_2)}$ for $\alpha<
-3(1+\omega_2)$, or $\rho_{_{DE}} \sim \tilde{C} a^\alpha$ for
$-3(1+\omega_2)<\alpha$, where $\tilde{C}=\frac{\alpha+3}{3
\omega_2(\alpha-\beta)}[(\beta+3) \rho_{_{DM0}}+(
\beta+3(1+\omega_2))(\rho_{_{DE0}}-C)]$. Thus, for $\alpha<
-3(1+\omega_2)$ we have $r=\frac{\rho_{_{DM}}}{\rho_{_{DE}}} \sim
\frac{C_1}{C} a^{\alpha+3(1+\omega_2)}$. Clearly in this case $r$
vanishes at late times, implying that the dark energy dominates over
the dark matter. For the second case we have that
\begin{eqnarray*}
r=\frac{\rho_{_{DM}}}{\rho_{_{DE}}} \sim
-\frac{\alpha+3(1+\omega_2)}{\alpha+3}=-1-\frac{3
\omega_2}{\alpha+3},
\end{eqnarray*}
thus alleviating the coincidence problem. In this case, in order to
have $r>0$ we must require $-3<\alpha<0$.

\subsection{Cosmologies with $\omega_2=0$}
Now we shall consider the generalization of
scenarios~(\ref{alphaRHO1A}) and~(\ref{alphaRHO2A}). This implies
that we must put $\omega_2=0$ and then Eq.~(\ref{Rho2}) describes
the behavior of the dark matter component, while the energy density
of the dark energy $\rho_{_{DE}}$ is described by Eq.~(\ref{Rho1}),
with $\omega_1<-1/3$. Thus the dark matter component has the
following form:
\begin{widetext}
\begin{eqnarray}\label{Rho2DM}
\rho_{_{DM}}=C a^{-3}+ \frac{(\alpha +3(1+\omega_1))(\beta
+3(1+\omega_1))\rho_{_{DE0}}}{3 \omega_1
(\beta-\alpha)}\left[a^\alpha-a^\beta \right]+
 \nonumber \\  \frac{C-\rho_{_{DM0}}}{3 \omega_1 (\beta-\alpha)} \left[
-(\beta+3)(\alpha +3(1+\omega_1))a^\alpha+(\alpha+3)(\beta
+3(1+\omega_1))a^{\beta} \right].
\end{eqnarray}
\end{widetext}
Notice that in general, as in the previous case, there are values of
the model parameters which lead to negative values of the dark
matter and/or the dark energy component (either at early or at late
times).

\begin{figure}
\includegraphics[width=8cm]{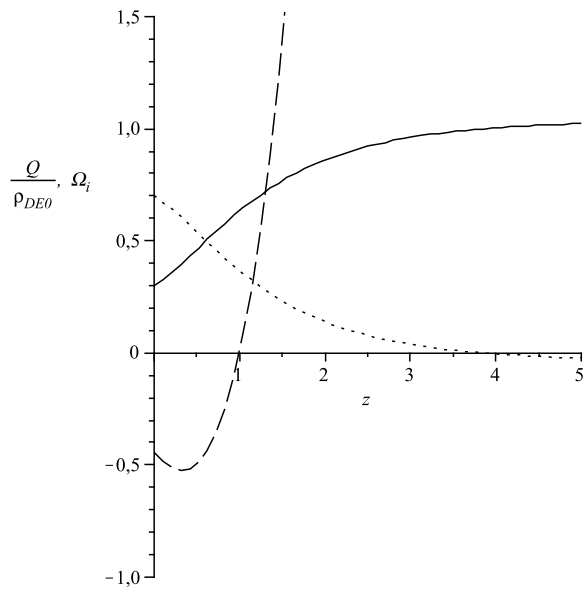}
\caption{\label{cero} We show the evolution of the dimensionless
energy densities $\Omega_i$ for dark matter (solid line), dark
energy (dotted line) and the corresponding behavior of the coupling
term $Q$ (dashed line) vs. redshift $z$ for the following values of
the model parameters: $k=0$, $\alpha=-2$, $\beta=-1$,
$\Omega_{_{DM0}}=0.3$, $\Omega_{_{DE0}}=0.7$, $\omega_1=-0.9$. In
this case the interacting term $Q$ is negative for $z<1$, implying
that for $z>1$ we have a transfer of energy from the dark matter to
the dark energy, and for $z<1$ the energy is being transferred from
the dark energy to the dark matter component.}
\end{figure}

In the following for the sake of simplicity we shall consider the
case $C=\rho_{_{DM0}}$. It is clear that this particular form of
Eqs.~(\ref{Rho1}) and~(\ref{Rho2DM}) contains as a particular case
the solution~(\ref{alphaRHO1A}) and~(\ref{alphaRHO2A}). This can be
seen directly by taking $\alpha=\tilde{\alpha}-3(1+\omega_1)$ and
$\beta=-3$ (this condition leads to $C_2=0$, $C_1=\rho_{_{DE0}}$).
In order for dark matter to drop during the expansion we shall
require that $\alpha<0$ and $\beta<0$. It is clear that for
$\omega_1 < -1/3$ there is a set of values of the model parameters
which leads to a positive energy density of dark matter during all
cosmic evolution. It can be shown that for $\alpha<-3$ and
$\beta<-3$ the energy density of the dark  matter takes negative
values at early times of the evolution. We may find a set of values
for the parameters which leads to a positive dark energy during all
evolution if $-3<\alpha<0$ and $-3<\beta<0$. For example in
Fig.~\ref{cero} we show a case where the energy density of dark
matter takes only positive values during all evolution. It is also
shown the behavior of the interacting term $Q$, which is positive at
the beginning of the expansion, becoming negative at some value of
the scale factor. Note that in Fig.~\ref{cero} is considered a flat
FRW scenario ($k=0$ and then $\Omega_{_{total}}=1$) with
$\Omega_{_{DM0}}=0.3$. This value of the dimensionless energy
density of the dark matter is in good agreement with a wide range of
observations: high-redshift Type Ia supernovae, evolution of
galactic clusters, high baryon content of clusters, lensing arcs in
clusters, and dynamical estimates from infrared galaxy
surveys~\cite{OBSDM,OBSDM1,OBSDM2}.
It is interesting to note that the interacting
term~(\ref{Qsolucion}) now is given by
\begin{eqnarray}
Q(a)=
\frac{(\alpha+3(1+\omega_1))(\beta+3(1+\omega_1))\rho_{_{10}}}{3\omega_1(\beta-\alpha)}
\left[-(\alpha+3)a^{\alpha-1}+(\beta+3)a^{\beta-1}\right].
\end{eqnarray}
Thus we have that it does not change sign only if $\alpha<-3$,
$\beta>-3$ or $\alpha>-3$, $\beta<-3$. However in this case there
are configurations leading to a negative energy density of the dark
matter component.

It is interesting to remark that, as in Sec. IV-A, for any solution
with $\omega_2=0$ we may find a set of cosmologies presenting a
scaling behavior for energy densities at early and late epochs.
Effectively, without any loss of generality let us suppose that
$\beta < \alpha <0$, and then at late times the dark energy
component will behave as $\rho_{_{DE}} \sim C_1 a^\alpha$ while the
dark matter component as $\rho_{_{DM}} \sim C a^{-3}$ for $\alpha<
-3$, or $\rho_{_{DM}} \sim \tilde{\tilde{C}} a^\alpha$ for
$\alpha>-3$, where $\tilde{\tilde{C}}=\frac{\alpha+3(1+\omega_1)}{3
\omega_1(\beta-\alpha)}[(\beta+3(1+\omega_1))
\rho_{_{DE0}}+(\rho_{_{DM0}}-C)( \beta+3)]$. Thus for $\alpha< -3$
we have for the ratio of energy densities
$r=\frac{\rho_{_{DM}}}{\rho_{_{DE}}} \sim \frac{C}{C_1}
a^{-(\alpha+3)}$. Clearly in this case $r$ diverges at late times,
implying that the dark matter component dominates over the dark
energy, so these models must be ruled out of consideration. On the
other hand, for $\alpha>-3$ we have cosmological models with a
scaling behavior since
\begin{eqnarray*}
r=\frac{\rho_{_{DM}}}{\rho_{_{DE}}} \sim
-\frac{\alpha+3(1+\omega_1)}{\alpha+3}=-1-\frac{3
\omega_1}{\alpha+3},
\end{eqnarray*}
thus alleviating the coincidence problem. In this case, in order to
have $r>0$ we must require $\alpha+3(1+\omega_1)<0$.

\subsection{The three-fluid interpretation}
We next consider models in which there are two dark matter sectors:
one describing the standard visible matter which is not interacting
with the dark energy component, and another which is interacting
with the dark energy. Both of these dark matter components are
treated as pressureless fluids. The advantage of such a description
is that we can describe the behavior of visible matter with an
energy density which decreases with the scale factor as $\rho_{m}
\sim a^{-3}$, while the energy density of the another dark matter
component behaves as $\rho_{_{DM}} \sim a^{n}$, with $n<0$ and $n
\neq -3$, in order that this second dark matter component also
decreases with the scale factor.

Let us consider the general solution~(\ref{Rho1}) and~(\ref{Rho2})
with $\omega_2=0$. In this case we shall not consider the energy
density $\rho_2$ in the normalized form~(\ref{Rho2DM}) and we shall
write it in the following form:
\begin{eqnarray}
\rho_2=\rho_{_{DM}}+\rho_m,
\end{eqnarray}
where
\begin{eqnarray}\label{AAA}
\rho_{_{DM}}=- \frac{\alpha +3(1+\omega_1)}{\alpha+3 }\, C_1
a^\alpha - \frac{\beta +3(1+\omega_1)}{\beta+3}\, C_2 a^\beta,
\nonumber \\ \rho_m=\rho_{m0} a^{-3}, \hspace{5.9cm}
\end{eqnarray}
and we have put $C=\rho_{m0}>0$. Note that, as in Sec. IV-B, the
energy density of the dark energy is given by $\rho_{_{DE}}:=\rho_1$
of Eq.~(\ref{Rho1}).

Thus, by putting expressions~(\ref{AAA}) and~(\ref{Rho1}) into the
conservation equation~(\ref{ConEq}), it may be rewritten as
\begin{eqnarray}
\dot{\rho}_{_{DM}}+\dot{\rho}_{_{DE}}+3H(\rho_{_{DM}}+(1+\omega_{_{DE}})\rho_{_{DE}})=0,
\nonumber \\  \dot{\rho}_{m}+3H\rho_m=0, \hspace{0.1cm}
\end{eqnarray}
where we have put $\omega_1:=\omega_{_{DE}}$. Clearly the
interpretation of these equations is different here: we have the
conservation of the total visible matter $\rho_m$, while the dark
matter is conserved together with the dark energy component.
Therefore we can associate an interaction between $\rho_{_{DM}}$ and
$\rho_{_{DE}}$. In this case the coupling between dark matter and
dark energy is still given by Eq.~(\ref{Qsolucion}).

\begin{figure}
\includegraphics[width=8cm]{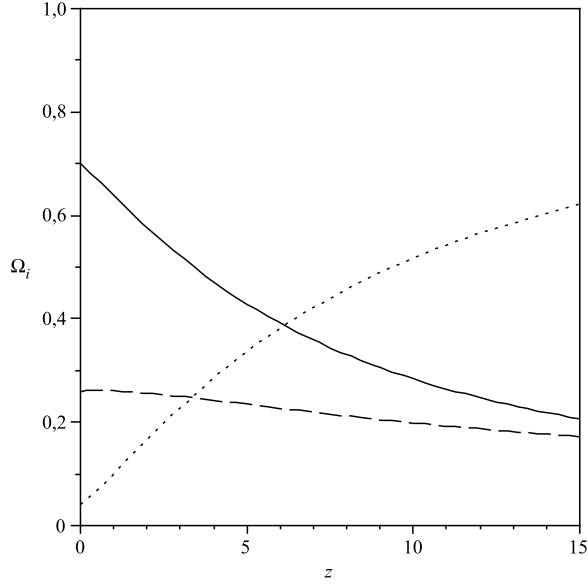}
\caption{\label{uno} We show the behavior of the dimensionless
energy densities $\Omega_i$ for standard visible matter (dotted
line), dark matter (dashed line) and dark energy (solid line) for
the following values of the model parameters: $k=0$,
$\Omega_{m0}=0.04$, $\Omega_{_{DM0}}=0.26$, $\omega_1=-2/3$. This
implies that $\beta< -1.542$ and $-1.542< \alpha <-1$. We have
plotted curves for $\alpha=-3/2$ and $\beta=-5/2$. For late times
the dark energy component dominates, while at early times dust is
dominating.}
\end{figure}

Normalizing the solution in order to have today
$\rho_{_{DM}}(a=1)=\rho_{_{DM0}}$ and
$\rho_{_{DE}}(a=1)=\rho_{_{DE0}}$ the constants $C_1$ and $C_2$ may
be written as
\begin{eqnarray}
C_1=\frac{(\alpha+3)\left[(\beta+3)\rho_{_{DM0}}+(\beta+3(1+\omega_1))\rho_{_{DE0}}\right]}{3
\omega_1(\alpha-\beta)}, \nonumber \\
C_2=-\frac{(\beta+3)\left[(\alpha+3)\rho_{_{DM0}}+(\alpha+3(1+\omega_1))\rho_{_{DE0}}\right]}{3
\omega_1(\alpha-\beta)}. \nonumber \\
\end{eqnarray}
It is clear that, if we want to have positive energy densities
during all cosmic evolution, we must require that all terms of
Eqs.~(\ref{Rho1}) and~(\ref{AAA}) must be positive. Thus, by
requiring $C_1 \geq 0$, $C_2 \geq 0$, $\frac{\alpha
+3(1+\omega_1)}{\alpha+3 }<0$, $\frac{\beta
+3(1+\omega_1)}{\beta+3}<0$ and without any lost of generality
$\beta < \alpha$ we have that the following constraints must be
satisfied for negative values of the state parameter $\omega_1$:
\begin{eqnarray*}
\frac{\beta
+3(1+\omega_1)}{\beta+3}  <  -\frac{\rho_{_{DM0}}}{\rho_{_{DE0}}}, \nonumber \\
-\frac{\rho_{_{DM0}}}{\rho_{_{DE0}}} < \frac{\alpha
+3(1+\omega_1)}{\alpha+3 } <0.
\end{eqnarray*}
This implies that the parameters $\alpha$ and $\beta$ are
constrained to be
\begin{eqnarray}\label{constraint15}
-\frac{3(1+r_0+\omega_1)}{1+r_0} < \alpha < -3(1+\omega_1),
\nonumber
\\
-3 < \beta < -\frac{3(1+r_0+\omega_1)}{1+r_0},
\end{eqnarray}
where $r_0=\frac{\rho_{_{DM0}}}{\rho_{_{DE0}}}$. It is clear that
these constraints imply that energy is being transferred from dark
energy to dark matter. Notice that for $C_1=0$ or $C_2=0$ we have
that the energy densities of dark matter and dark energy are
functions of the same power of the scale factor, i.e.
$\rho_{_{DE}}/\rho_{_{DM}}=const$.

\begin{figure}
\includegraphics[width=8cm]{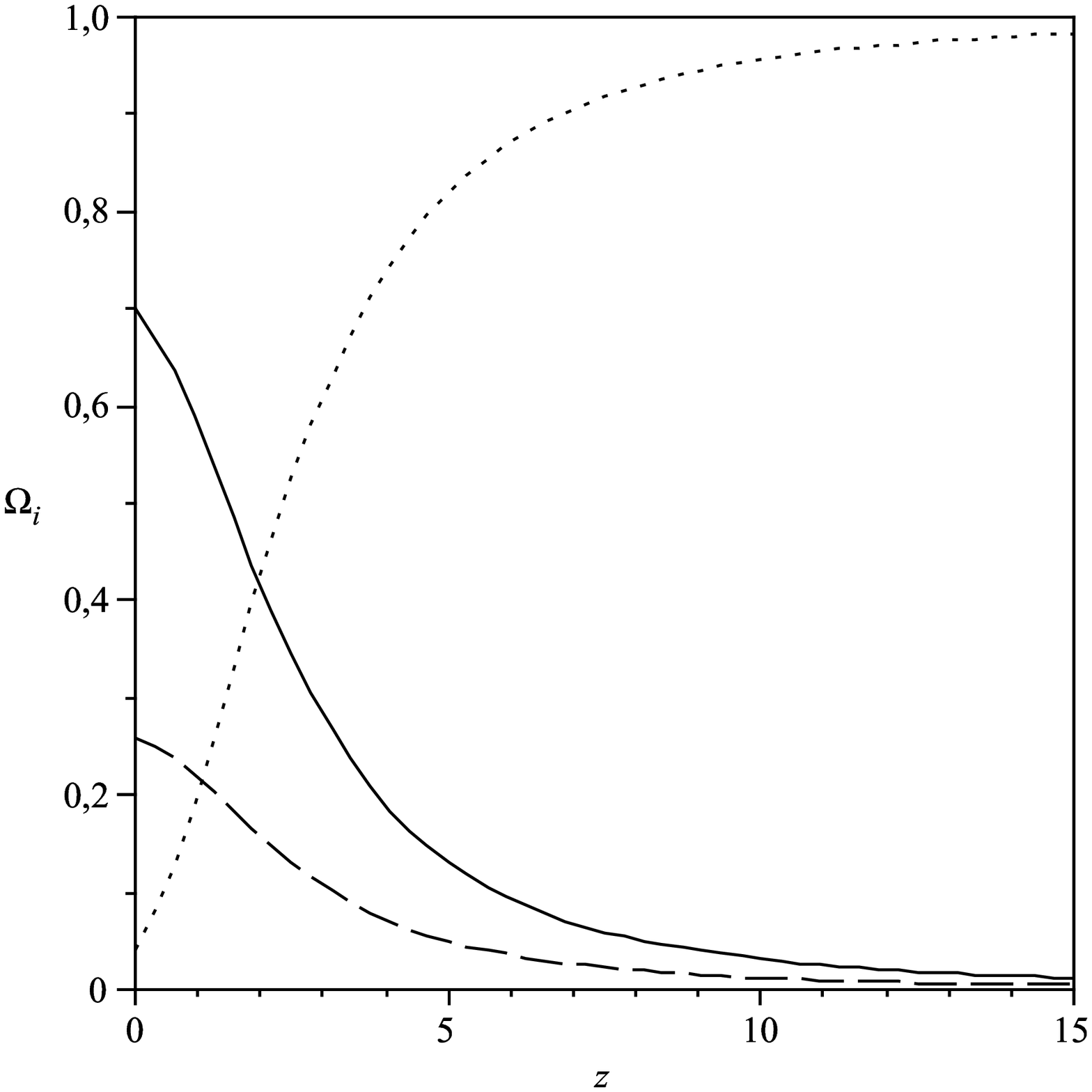}
\caption{\label{dos} We show the behavior of the dimensionless
energy densities $\Omega_i$ for standard visible matter (dotted
line), dark matter (dashed line) and dark energy (solid line) for
the following values of the model parameters: $k=0$,
$\Omega_{m0}=0.04$, $\Omega_{_{DM0}}=0.26$, $\omega_1=-1.2$. This
implies that $\beta< -0.375$ and $-0.375< \alpha <0$ and we have
plotted curves for $\alpha=-0.3$ and $\beta=-1/2$. For late times
the dark energy component dominates, while at early times dust is
dominating. Note that in this case the dust-dark energy equality and
dust-dark matter equality occur for $z\thickapprox 2$ and $z
\thickapprox 1$ respectively.}
\end{figure}

\begin{figure}
\includegraphics[width=8cm]{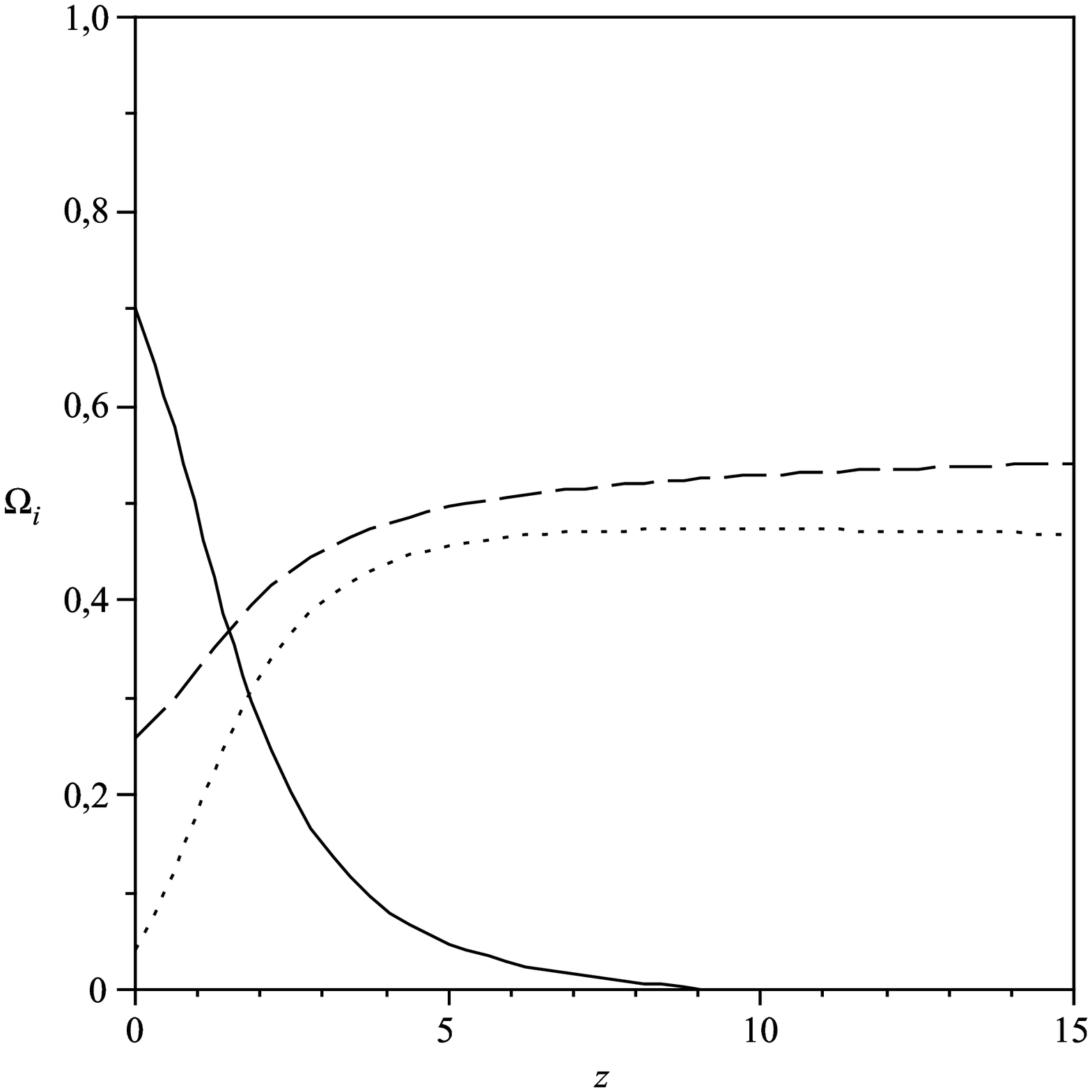}
\caption{\label{tres} We show the behavior of the dimensionless
energy densities $\Omega_i$ for standard visible matter (dotted
line), dark matter (dashed line) and dark energy (solid line) for
$k=0$, $\Omega_{m0}=0.04$, $\Omega_{_{DM0}}=0.26$, $\omega_1=-1.2$,
$\alpha=-0.275$ and $\beta=-3.1$. In this case the dark energy
density becomes negative for $z \gtrsim 9$ and the dust-dark energy
equality occurs for $z\thickapprox 1.5$, while for $z \thickapprox
2$ we have the dark energy-dark matter equality. Notice that for $z
\gtrsim 5$ the energy is transferred from the dark matter to dark
energy, and for $z \lesssim 5$ is from dark energy to dark matter.}
\end{figure}

In Figs.~\ref{uno} and~\ref{dos} are plotted the evolution of three
dimensionless energy densities $\Omega_{m}$, $\Omega_{_{DM}}$ and
$\Omega_{_{DE}}$ ($\Omega_{m}+\Omega_{_{DM}}+\Omega_{_{DE}}=1$) in
the case of flat FRW cosmologies ($k=0$) and always positive dark
energy and dark matter energy densities. In order to have an
estimation for the behavior of dimensionless energy densities, the
cosmological model parameters are fixed to take the values
$\Omega_{_{m0}}=0.04$ and $\Omega_{_{DM0}}=0.26$, which are in good
agreement with a wide range of
observations~\cite{OBSDM,OBSDM1,OBSDM2}. In both cases the energy is
transferred from the dark energy to the dark matter.

If we do not exclude the possibility of negative energies we can
also consider cases where the dark energy becomes negative at some
stage during the cosmic evolution. For example in Fig.~\ref{tres} we
show the case where negative values for the dark energy density are
allowed at early times. However it must be noticed that the total
energy density $\rho_{_{tot}}=\rho_{m0}+\rho_{_{DM0}}+\rho_{_{DE0}}$
is positive during all cosmic evolution. An interesting feature of
such a scenario with negative energy is that we have a period where
the energy is transferred from dark energy to dark matter and
another where this occurs from the dark matter component to dark
energy.

\section{conclusions}
In this paper we have studied Friedmann-Robertson-Walker
cosmological models with matter content composed of two barotropic
perfect fluids where one of the energy densities is given by a sum
of two powers of the scale factor of the form of Eq.~(\ref{Rho1}).
By associating with these cosmologies an interacting term $Q$, it
can be shown that interacting scenarios with couplings given by
$Q=\tilde{\alpha} H \rho_1$, $Q=\tilde{\alpha} H \rho_2$,
$Q=\tilde{\alpha} H (\rho_1+ \rho_2)$ and $Q=\tilde{\alpha} H
\rho_1+\tilde{\beta} H \rho_2$ (with constants $\tilde{\alpha}$ and
$\tilde{\beta}$) correspond to particular cases of our cosmological
model. The studied cosmological models contain a class of solutions
having a scaling behavior at early and at late times, and then the
coincidence problem is substantially alleviated.

It is interesting to note that in the framework of the considered
scenarios it is possible to introduce a three fluid interpretation,
where one fluid describes the standard visible matter and is
conserved separately from the dark matter and dark energy
components, which are conserved together. If we want to introduce
the interacting picture, thus the visible matter is not interacting
neither with the dark matter nor dark energy components, while these
latter two components are interacting with each other. In this case,
if energy densities of dark matter and dark energy are positive
during all cosmic evolution, then the energy always is transferred
from the dark energy component to dark matter in agreement with the
conclusions of Ref.~\cite{Pavon2007}.

It is remarkable that the inequalities~(\ref{constraint15}) impose a
lower limit on the values of the state parameter $\omega_1$.
Effectively, if we require that the energy density of the dark
matter be decreasing with the expansion, i.e. $\alpha<0$ and
$\beta<0$, then the state parameter is constrained to be in the
range $\omega_1>-(1+r_0)$. Thus, in agreement with the observations,
we have for the ratio
$r_0=\frac{\rho_{_{DM0}}}{\rho_{_{DE0}}}=\frac{\Omega_{_{DM0}}}{\Omega_{_{DE0}}}=0.26/0.7$
and then $\omega_1>-1.371$. Note that this constraint on the state
parameter $\omega_1$ is allowed in this model by the requirement of
positivity of energy densities of dark matter and dark energy during
all cosmic evolution. However, one can impose more stringent
constraints on the model parameters by using for example the latest
supernova data. In order to do this we shall use the recent
compilation of SNe Ia data called the Union data
set~\cite{Kowalski}, which contains a set of 57 nearby ($0.015 < z <
0.15$) Type Ia supernovae, and 250 high-redshift supernovae. The
$\chi^2$ statistic is quite helpful in constraining the parameter
values of a given model~\cite{LVerde}. In our case this method will
allow us to fit the set of the model parameters $\alpha$, $\beta$
and $\omega_1$ to the set of cosmological parameters of the Union
compilation data, by finding the best fit values of the model
parameters by minimizing this $\chi^2$. In this case the minimum of
$\chi^2$ should be roughly equal to the number of data, or the
so-called ``reduced chisquare" $\chi^2_{\nu}=\chi^2_{min}/307$
should be roughly equal to 1.

\begin{figure}
\includegraphics[width=8cm]{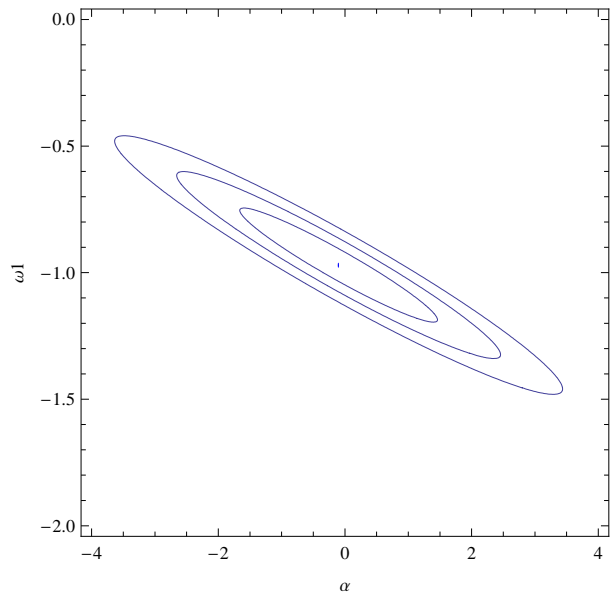}
\caption{\label{Fabuno} We show the probability contours at $68.3
\%$, $95.4 \%$ and $99.7 \%$ confidence levels in the
$\alpha-\omega_1$ plane for the value $\beta=-5/2$ of
Fig.~\ref{uno}. In this case the best-fitting parameters are
$\alpha=-0.1$ and $\omega_1=-0.97$ with $\chi^2_{\nu}=1.021$.}
\end{figure}

In Fig.~\ref{Fabuno} we show the probability contours from the above
$307$ data points only at $68.3 \%$, $95.4 \%$ and $99.7 \%$
confidence levels (from inside to outside) in the $\alpha-\omega_1$
plane for the value $\beta=-5/2$ of Fig.~\ref{uno}. In this case the
best-fitting parameters are $\alpha=-0.1$ and $\omega_1=-0.97$ with
$\chi^2_{\nu}=1.021$, implying the constraints $-3.6 \lesssim \alpha
\lesssim 3.4$ and $-1.49 \lesssim \omega_1 \lesssim -0.45$ ($99.7
\%$ C.L.). Note that in this case, by taking into account the above
constraint on the state parameter $\omega_1$, we can rewrite the
latter constraint as $-1.371<\omega_1 \lesssim -0.45$.

\begin{figure}
\includegraphics[width=8cm]{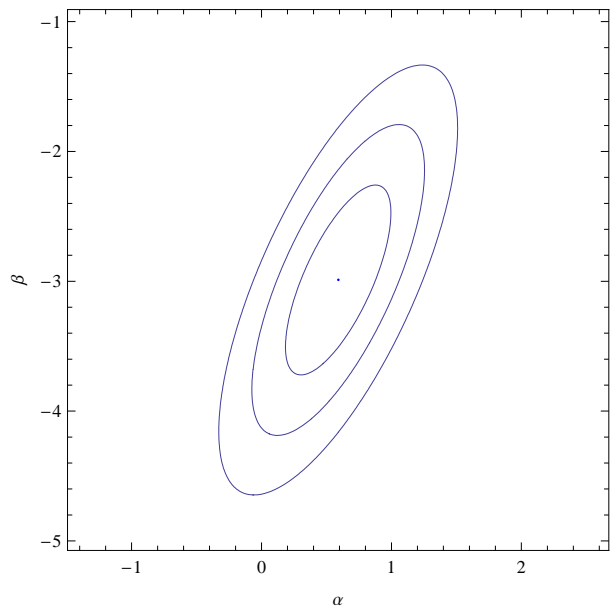}
\caption{\label{FabunoA} We show the probability contours at $68.3
\%$, $95.4 \%$ and $99.7 \%$ confidence levels (from inside to
outside) in the $\alpha-\beta$ plane for the value $\omega_1=-1.2$
of Fig.~\ref{uno}. In this case the best-fitting parameters are
$\alpha=0.59$ and $\beta=-2.99$ with $\chi^2_{\nu}=1.024$.}
\end{figure}

In Fig.~\ref{FabunoA} we show the probability contours at $68.3 \%$,
$95.4 \%$ and $99.7 \%$ confidence levels (from inside to outside)
in the $\alpha-\beta$ plane for the value $\omega_1=-1.2$ of
Fig.~\ref{dos} and Fig.~\ref{tres}. In this case the best-fitting
parameters are $\alpha=0.59$ and $\beta=-2.99$ with
$\chi^2_{\nu}=1.024$, implying the constraints $-0.34 \lesssim
\alpha \lesssim 1.52$ and $-4.65 \lesssim \beta \lesssim -1.33$
($99.7 \%$ C.L.). Thus, in this case, in order to have a decreasing
with expansion dark matter energy density, we must require that $-3<
\beta \lesssim -1.33$ and $-0.34 \lesssim \alpha <0$.

\begin{figure}
\includegraphics[width=8cm]{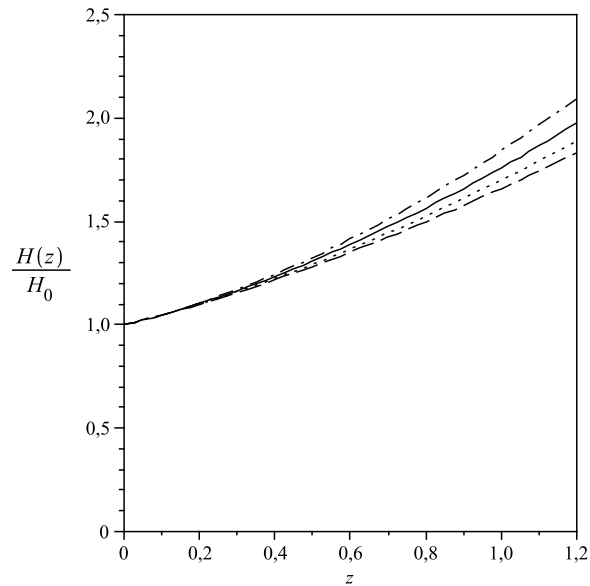}
\caption{\label{siete} We show the evolution of the Hubble parameter
with the redshift $z$ for $k=0$, $\Omega_{m0}=0.04$,
$\Omega_{_{DM0}}=0.26$, $\beta=-0.05$, $\omega_1=-1$ and
$\alpha=-2.6$ (dashed line), $\alpha=-2.8$ (dotted line) and
$\alpha=-3.4$ (dashdotted line). For comparison, the prediction of
the $\Lambda$CDM model (solid line) is also shown.}
\end{figure}

\begin{figure}
\includegraphics[width=8cm]{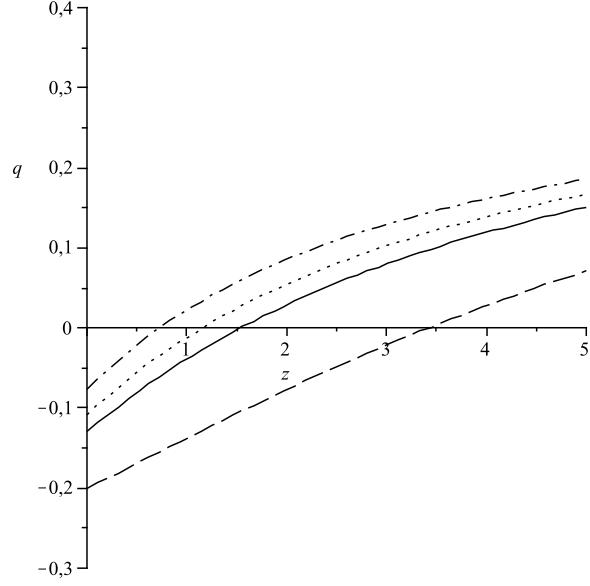}
\caption{\label{cuatro} We show the evolution of the deceleration
parameter $q$ with the redshift $z$ for $k=0$, $\Omega_{m0}=0.04$,
$\Omega_{_{DM0}}=0.26$, $\alpha=-3/2$ and $\beta=-5/2$, and
$\omega_1=-0.55$ (dashdotted line), $\omega_1=-0.58$ (dotted line),
$\omega_1=-0.6$ (solid line) and $\omega_1=-2/3$ (dashed line). In
this case the value of the redshift $z$, for transition from
decelerated expansion to accelerated expansion, becomes smaller with
the increase of the state parameter $\omega_1$.}
\end{figure}

\begin{figure}
\includegraphics[width=8cm]{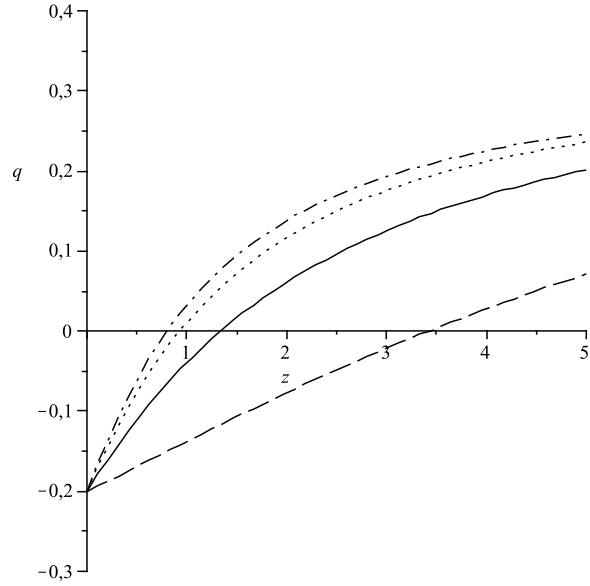}
\caption{\label{cinco} We show the evolution of the deceleration
parameter $q$ with the redshift $z$ for $k=0$, $\Omega_{m0}=0.04$,
$\Omega_{_{DM0}}=0.26$, $\beta=-5/2$, $\omega_1=-2/3$, and
$\alpha=-3/2$ (dashed line), $\alpha=-1.2$  (solid line),
$\alpha=-1$ (dotted line) and $\alpha=-0.9$  (dashdotted line). In
this case the value of the redshift $z$, for transition from
decelerated expansion to accelerated expansion, becomes smaller with
the increase of the parameter $\alpha$.}
\end{figure}

\begin{figure}
\includegraphics[width=8cm]{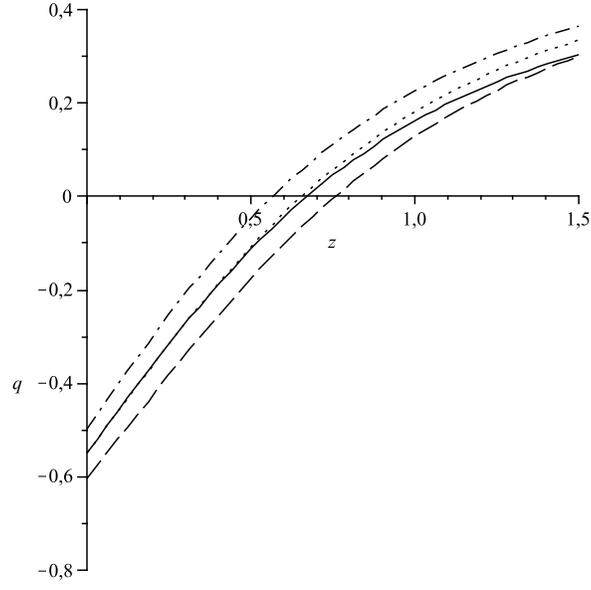}
\caption{\label{seis} We show the evolution of the deceleration
parameter $q$ with the redshift $z$ for $k=0$, $\Omega_{m0}=0.04$,
$\Omega_{_{DM0}}=0.26$, $\alpha=-3.1$, $\beta=-0.05$, and
$\omega_1=-1.05$ (dashed line), $\omega_1=-1$ (dotted line) and
$\omega_1=-0.95$ (dashdotted line). For comparison, the prediction
of the $\Lambda$CDM model (solid line) is also shown.}
\end{figure}

Lastly, in Fig.~\ref{siete} we show the evolution of the Hubble
parameter $H(z)/H_0$ vs. the redshift $z$ for $\omega_1=-1$, while
in Figs.~\ref{cuatro}, \ref{cinco} and~\ref{seis} we show the
evolution of the deceleration parameter $q$ vs. redshift $z$ for
some values of the state parameter $\omega_1$
($-1.371<\omega_1<-1/3$). Note that in Fig.~\ref{seis}, for
comparison, the prediction of the $\Lambda$CDM model is also shown.

\section{Acknowledgements}
This work was supported by CONICYT through Grant FONDECYT N$^0$
1080530 (MC), PhD Grant N$^0$ 21070949 (FA) and by Direcci\'on de
Investigaci\'on de la Universidad del B\'\i o--B\'\i o (MC).

\end{document}